\documentstyle[aps,prd,epsfig,preprint]{revtex}
\def\beq{\begin{equation}}
\def\eeq{\end{equation}}
\def\bea{\begin{eqnarray}}
\def\eea{\end{eqnarray}}

\def\fun#1#2{\lower3.6pt
\vbox{\baselineskip0pt\lineskip.9pt
\ialign{$\mathsurround=0pt#1\hfill##\hfil$
\crcr#2\crcr\sim\crcr}}}
\begin{document}
\vspace{0.5in}
\title{\vskip-2.5truecm
{\hfill \baselineskip 14pt {\hfill {{\small \hfill 
UT-STPD-3/00}}} \\ 
{{\small \hfill FTUAM 00-09}} 
\vskip .1truecm} 
\vspace{1.0cm} 
\vskip 0.1truecm
{\bf Cold Dark Matter and $b\rightarrow s\gamma$ in the 
Ho\v{r}ava-Witten Theory }} 
\vspace{1cm}
\author{{S. Khalil}$^{(1),(2)}$\thanks{shaaban.khalil@uam.es},
{G. Lazarides}$^{(3)}$\thanks{lazaride@eng.auth.gr} 
{and  C. Pallis}$^{(3)}$\thanks{kpallis@gen.auth.gr}} 
\vspace{1.0cm}
\address{$^{(1)}${\it Departmento de Fisica
Te\'orica, C.XI, Universidad Aut\'onoma de Madrid,\\ 28049
Cantoblanco, Madrid, Spain.}}
\address{$^{(2)}${\it Ain Shams University, Faculty of
Science, Cairo 11566, Egypt.}}
\address{$^{(3)}${\it Physics Division, School of Technology,
Aristotle University of Thessaloniki,\\ Thessaloniki GR 540 06,
Greece.}}
\maketitle 

\vspace{1cm}

\begin{abstract}
\baselineskip 12pt

\par
The minimal supersymmetric standard model with complete, partial 
or no Yukawa unification and radiative electroweak breaking with 
boundary conditions from the Ho\v{r}ava-Witten theory is 
considered. The parameters are restricted by constraining the 
lightest sparticle relic abundance by cold dark matter 
considerations and requiring the $b$-quark mass after 
supersymmetric corrections and the branching ratio of 
$b\rightarrow s\gamma$ to be compatible with data. Complete 
Yukawa unification can be excluded. Also, $t-b$ Yukawa 
unification is strongly disfavored since it requires almost 
degenerate lightest and next-to-lightest sparticle masses. 
However, the $b-\tau$ or no Yukawa unification cases avoid 
this degeneracy. The latter with $\mu<0$ is the most natural 
case. The lightest sparticle mass, in this case, can be as low 
as about $77~{\rm GeV}$.

\end{abstract}

\thispagestyle{empty}
\newpage
\pagestyle{plain} \setcounter{page}{1} \baselineskip 20pt

\par
Recently, it has been realized that the five existing perturbative 
string theories (type I open strings, type IIA and IIB closed 
strings, and the $E_8 \times E'_8$ and $SO(32)$ closed heterotic 
strings) and the 11-dimensional supergravity correspond to 
different vacua of a unique underlying theory, called M-theory. 
Ho\v{r}ava and Witten have shown \cite{witten1} that the strong 
coupling limit of the $E_8 \times E'_8$ heterotic string theory 
is equivalent to the low energy limit of M-theory compactified on 
$S^1/Z_2$ which is a line segment of length $\rho$. As 
$\rho\to 0$, the weakly coupled heterotic string is recovered. 
The observable $E_8$ gauge fields reside in one (10-dimensional) 
end of this segment, while the hidden sector $E'_8$ gauge fields 
reside in its other end. Gravitational fields propagate in the 
11-dimensional bulk. 

\par
The main success of the Ho\v{r}ava-Witten theory is that it solves, 
in an elegant way, the gauge coupling unification problem, i.e., the 
discrepancy between the supersymmetric (SUSY) grand unified theory 
(GUT) scale $M_X\simeq 2\times 10^{16}{\rm{GeV}}$ (consistent 
with the data on the low energy gauge coupling constants) and the 
string unification scale 
$M_{str}\simeq 5\times 10^{17}{\rm{GeV}}$  
calculated in the weakly coupled string theory. Before M-theory, 
there were several proposals (such as large threshold corrections, 
intermediate scales, and extra particles) for explaining this 
discrepancy but none was totally satisfactory. In the 
strongly coupled heterotic string theory, the extra Kaluza-Klein 
states do not affect the running of the gauge coupling constants, 
which live on the boundary of the 11-dimensional spacetime. On 
the contrary, they accelerate the running of the gravitational 
coupling constant and, thus, reduce $M_{str}$ to $M_{X}$. 
Moreover, SUSY breaking in M-theory naturally leads 
\cite{nilles} to gaugino masses of the order of the gravitino 
mass in contrast to the weakly coupled heterotic string case where 
the gaugino masses were tiny. 

\par
Similarly to the weakly coupled heterotic string, the
compactification of the Ho\v{r}ava-Witten theory can lead to 
the spontaneous breaking of $E_8$ to phenomenologically more 
interesting groups. The simplest breaking of $E_8$ to $E_6$ is 
achieved \cite{witten2} by the so-called standard embedding 
(SE), where the holonomy group of the spin connection of a 
Calabi-Yau three-fold is identified with a $SU(3)$ subgroup of 
$E_8$. Further breaking of $E_6$ to semi-simple groups such as 
the trinification group $SU(3)_c\times SU(3)_L\times SU(3)_R$ 
and the flipped $SU(6)\times U(1)$ group can be performed 
via Wilson loops. The trinification group contains $SU(2)_R$. 
Assuming then that the Higgs doublets and the third family 
right-handed quarks form $SU(2)_R$ doublets, one obtains 
\cite{pana} the `asymptotic' Yukawa coupling relation $h_t=h_b$ 
and, hence, large $\tan\beta\approx m_{t}/m_{b}$. The flipped 
$SU(6)$, for certain embeddings of the minimal supersymmetric 
standard model (MSSM) fields, contains \cite{pana1} $SU(4)_c$. 
Requiring that the third family lepton doublet belongs to $SU(6)$ 
15-plets and the right-handed $b$-quark as well as the Higgs doublet 
coupling to the down-type quarks belong to $SU(6)$ $\bar 6$-plets, 
one gets `asymptotic' $b-\tau$ Yukawa unification ($h_b=h_\tau$).

\par
In the strongly coupled case, the SE is not special 
\cite{ovrut}. Non-standard embeddings (NSE) may 
lead to simple gauge groups such as $SU(5)$ or $SO(10)$ which 
could yield $b-\tau$ or complete ($h_t=h_b=h_\tau$) Yukawa 
unification. However, in general, we do not obtain 
Higgs superfields in the adjoint representation. Further gauge 
symmetry breaking then requires Wilson loops and, thus, (partial) 
Yukawa unification is lost. This may be avoided by employing 
special constructions with higher Kac-Moody level \cite{ibanez}. 
Complete Yukawa unification can be obtained in the Pati-Salam  
gauge group $SU(4)_c\times SU(2)_L\times SU(2)_R$ which may 
arise in NSE. This group contains both $SU(4)_c$ and $SU(2)_R$ 
and does not require Wilson loops for its breaking. Furthermore, 
in string theories where the couplings have a common origin, 
partial or complete Yukawa unification can be realized even 
without a unified gauge group \cite{shaaban}. Thus all four 
possibilities with complete, partial ($t-b$ or $b-\tau$) or 
no Yukawa unification are in principle allowed.

\par
The soft SUSY breaking in the SE and NSE cases has been 
studied in Ref.\cite{km}. One obtains universal 
boundary conditions, i.e., a common scalar mass $m_0$, a common 
gaugino mass $M_{1/2}$ and a common trilinear coupling $A_0$ 
given by (with zero vacuum energy density and no CP 
violating phases)
\begin{equation}
m_0^2=m_{3/2}^2-\frac{3m_{3/2}^2}{(3+\epsilon)^2}\left(
\epsilon(6+\epsilon)\sin^2\theta+(3+2\epsilon)\cos^2\theta
-2\sqrt{3}\epsilon\cos\theta\sin\theta\right), 
\label{m0}
\end{equation}
\begin{equation}
M_{1/2}=\frac{\sqrt{3}m_{3/2}}{1+\epsilon}(\sin\theta
+\frac{\epsilon}{\sqrt{3}}\cos\theta ), 
\label{Mgaug}
\end{equation}
\begin{equation}
A_0=-\frac{\sqrt{3}m_{3/2}}{3+\epsilon}\left(
(3-2\epsilon)\sin\theta+\sqrt{3}\epsilon\cos\theta\right),
\label{A0}
\end{equation}
where $m_{3/2}$ is the gravitino mass, $\theta$ 
($0<\theta<\pi/2$) is the goldstino angle, and the parameter 
$\epsilon$ lies between 0 ($-1$) and 1 in the SE (NSE) case 
\cite{km}. The range of $\epsilon$ is the only difference 
between the two embeddings at the level of soft SUSY breaking.

\par
In this paper, we will study the MSSM which results from the 
Ho\v{r}ava-Witten theory. We will assume radiative electroweak
symmetry breaking with the universal boundary conditions  
in Eqs.(\ref{m0})-(\ref{A0}) and examine all cases with
complete, partial ($t-b$ or $b-\tau$) or no Yukawa unification.
Our main aim is to restrict the parameter space by 
simultaneously imposing a number of phenomenological and 
cosmological constraints. In particular, the $b$-quark mass
after including SUSY corrections and the branching ratio of 
$b\rightarrow s\gamma$ should be compatible with data. Also, the 
lightest supersymmetric particle (LSP) is required to provide 
the cold dark matter (CDM) in the universe. Its relic abundance 
must then be consistent with either of the two available 
cosmological models with zero/nonzero cosmological constant, 
which provide the best fits to all the data (see 
Refs.\cite{cdm,lahanas}).

\par
The GUT scale $M_X$ and gauge coupling constant 
are determined by using the 2-loop SUSY renormalization group 
equations (RGEs) for the gauge and Yukawa coupling constants 
between $M_X$ and a common SUSY threshold 
$M_S\approx\sqrt{m_{\tilde t_1}m_{\tilde t_2}}$ 
($\tilde t_{1,2}$ are the stop quark mass eigenstates), 
which minimizes the radiative corrections to $\mu$ and $m_A$ 
(see e.g., Ref.\cite{cd2}). Between $M_S$ and $m_Z$, we 
take the standard model (SM) 1-loop RGEs. The $t$-quark 
and $\tau$-lepton masses are fixed to their central 
experimental values $m_t(m_t)=166~{\rm{GeV}}$ and 
$m_\tau(m_\tau)=1.78~{\rm{GeV}}$.
The asymptotic values of $h_t$, $h_\tau$ are then determined 
for each $\tan\beta$ at $M_S$ and $h_b$ is derived from 
$t-b$ or $b-\tau$ Yukawa unification. The resulting $m_b(m_Z)$ 
is compared to its experimental value 
$m_b(m_Z)\simeq 2.67\pm 0.98~{\rm{GeV}}$ \cite{mb} 
(with a $95\%$ confidence margin) after 
1-loop SUSY corrections. For complete Yukawa 
unification, $\tan\beta$ at $M_S$ is fixed. For no Yukawa 
unification, $h_b$ is adjusted so that the corrected 
$m_b(m_Z)=2.67~{\rm{GeV}}$. $M_S$ is 
specified consistently with the SUSY spectrum.

\par
We next integrate the 1-loop RGEs for the soft SUSY 
breaking terms assuming universal boundary conditions given by 
Eqs.(\ref{m0})-(\ref{A0}). At $M_S$, we impose the 
minimization conditions to the tree-level renormalization group 
improved potential and calculate the Higgsino mass $\mu$ 
(up to its sign). The sparticle spectrum is evaluated at 
$M_S$. The LSP, which is the lightest neutralino 
($\tilde\chi$), turns out to be bino-like with purity $>98\%$ 
for almost all values of the parameters. The next-to-lightest 
supersymmetric particle (NLSP) is the lightest stau 
($\tilde\tau_2$). Since we consider large  
$\tan\beta$'s too, we are obliged to include the third 
generation sfermion mixing. The mixing of the lighter generation 
sfermions, however, remains negligible due to the small masses 
of the corresponding fermions. Furthermore, we take into account 
the 2-loop radiative corrections \cite{fh} to the CP-even 
neutral Higgs boson masses $m_h$, $m_H$, which turn out to be 
sizeable for the lightest boson $h$.

\par
Our calculation depends on the following free parameters: 
${\rm sign}\mu$, $\tan\beta$, $m_{3/2}$, $\epsilon$, 
$\theta$. The relation found in Ref.\cite{copw} between the 
CP-odd Higgs boson mass $m_A$ and the asymptotic scalar and 
gaugino masses, takes, in our case, the form 
\begin{equation}
m^2_A \simeq c_{3/2}m_{3/2}^2+c_s m_{3/2}^2\sin^2\theta
+c_{2s}m_{3/2}^2\sin{2\theta}-m_Z^2 ~, 
\label{mAc}
\end{equation}
where the coefficients $c_{3/2}\sim 0.1$, $c_s, c_{2s}\sim 1$ 
depend on $\tan\beta$, $\epsilon$, and 
$M_S$. We verified that this relation holds with an accuracy 
better than $0.02\%$. We use it to express $m_{3/2}$ in terms 
of $m_A$ for fixed ${\rm sign}\mu$, $\tan\beta$, $\epsilon$ 
and $\theta$ ($M_S$ is determined self-consistently from the 
SUSY spectrum). The free parameter $m_{3/2}$ can, thus, be 
replaced by $m_A$.

\par
In practice, the number of free parameters can be reduced by one. To 
see this, we fix ${\rm sign}\mu$, $\tan\beta$ and $m_A$ and 
observe that, along the lines in the $\epsilon-\theta$ plane 
where $m_0$ and $M_{1/2}$ remain constant, $A_0$ varies only by 
a few per cent. Consequently, the whole sparticle spectrum (except 
the gravitino mass) remains essentially unchanged along these lines 
which we call equispectral lines. Thus, for all practical purposes, 
$\epsilon$ and $\theta$ can be replaced by a single parameter 
which we choose to be the relative mass splitting between the LSP 
and the NLSP $\Delta_{NLSP}=
(m_{\tilde\tau_2}-m_{\tilde\chi})/m_{\tilde\chi}$.
Our final free parameters then are ${\rm sign}\mu$, $\tan\beta$, 
$m_A$, $\Delta_{NLSP}$.
Note that, for fixed $\epsilon$, $\Delta_{NLSP}$ increases as 
$\theta$ decreases. Also, for fixed $\theta>\pi/6~(<\pi/6)$, 
$\Delta_{NLSP}$ decreases (increases) as $\epsilon$ increases. 
Finally, we find that $\Delta_{NLSP}$ is maximized, generally, 
at $\theta=\pi/9$ and $\epsilon\to 1$. Our calculation is 
performed at an appropriate value of $\epsilon$ in each case so 
that all relevant $\Delta_{NLSP}$'s can be obtained.

\par
An important constraint results from the inclusive branching ratio 
of $b\rightarrow s\gamma$ \cite{bsg}, which is calculated here 
by using the formalism of Ref.\cite{kagan}. The dominant 
contributions, besides the SM one, come from the charged Higgs 
bosons ($H^\pm$) and the charginos. The former interferes 
constructively with the SM contribution, while the latter 
interferes constructively (destructively) with the other two 
contributions when $\mu>0$ ($\mu<0$). The SM contribution, 
which is factorized out in the formalism of Ref.\cite{kagan}, 
includes the next-to-leading order (NLO) QCD \cite{nlosm} and 
the leading order (LO) QED \cite{kagan,loqed} corrections. The 
NLO QCD corrections \cite{nlohiggs} to the charged Higgs boson 
contribution are taken from the first paper in Ref.\cite{nlohiggs}. 
The SUSY contribution is evaluated by including only the LO QCQ 
corrections using the formulae in Ref.\cite{nlosusy}. 
NLO QCD corrections to the SUSY contribution have also been 
discussed in Ref.\cite{nlosusy}, but only under certain very 
restrictive conditions which never hold in our case since the  
chargino and lightest stop quark masses are comparable to the 
masses of the other squarks and the gluinos. We, thus, do not 
include these corrections in our calculation. 

\par
The branching ratio ${\rm BR}(b\rightarrow s\gamma)$ is 
first evaluated with central values of the input parameters 
and the renormalization and matching scales. We find that, 
for each ${\rm sign}\mu$, $\tan\beta$ and $\Delta_{NLSP}$, 
there exists a value of $m_A$ above which the 
${\rm BR}(b\rightarrow s\gamma)$ enters and remains in the 
experimentally allowed region \cite{cleo}: $2\times 10^{-4}
\stackrel{_{<}}{_{\sim }}{\rm BR}(b\rightarrow s\gamma)
\stackrel{_{<}}{_{\sim }}4.5 \times 10^{-4}$. This lower 
bound on $m_A$ corresponds to the upper (lower) bound on the 
branching ratio for $\mu>0 $ ($\mu<0 $) and, for most of the 
parameter space, is its absolute minimum. For relatively small 
$\tan\beta$'s, however, the absolute minimum of $m_A$ comes 
from the experimental bound 
$m_h\stackrel{_{>}}{_{\sim }}113.4~{\rm GeV}$. We take 
$\tan\beta\gtrsim 2.3$ since otherwise $m_h$ is too small.

\par
The lower bound on $m_A$ can be considerably reduced if the 
theoretical uncertainties entering into the calculation of 
${\rm BR}(b\rightarrow s\gamma)$ are taken into account. 
These uncertainties originating from the experimental errors 
in the input parameters and the ambiguities in the 
renormalization and matching scales are known to be quite 
significant. The SM and charged Higgs contributions generate 
an uncertainty of about $\pm 10\%$ (see first paper in 
Ref.\cite{nlohiggs}). The uncertainty from the SUSY 
contribution cannot be reliably calculated at the moment 
since the NLO QCD corrections to this contribution are not 
known in our case. Fortunately, the SUSY contribution is 
pretty small in all cases which are crucial for our 
qualitative conclusions. Be that as it may, we take the 
uncertainty from this contribution, evaluated at the LO in 
QCD, to be about $\pm 30\%$. 

\par
For large or intermediate $\tan\beta$'s, a severe restriction 
arises from the sizable SUSY corrections to the $b$-quark mass. 
The dominant contributions are from the sbottom-gluino and 
stop-chargino loops and are calculated by using the simplified 
formulae of Ref.\cite{pierce}. We find here that the size of 
these corrections practically depends only on $\tan\beta$ 
(compare with Refs.\cite{cd2,copw}). Also, their sign is 
opposite to the one of $\mu$ in contrast to the chargino 
contribution to the ${\rm BR}(b\rightarrow s\gamma)$ which, 
as mentioned, has the sign of $\mu$.

\par
An additional restriction comes from the LSP cosmic relic abundance. 
We calculate this abundance by closely following the formalism of 
Ref.\cite{cdm} where $\tilde\chi-\tilde\tau_2$ coannihilations 
\cite{ellis} have been consistently included for all values of 
$\tan\beta$. However, coannihilations \cite{coan} of these 
sparticles with the lighter generation right-handed sleptons 
$\tilde e_R$, $\tilde e_R^\ast$, $\tilde\mu_R$, 
$\tilde\mu_R^\ast$ (considered degenerate), which were ignored in 
Ref.\cite{cdm}, are now important and must be included since our 
calculation here extends to small ($\lesssim 15$) $\tan\beta$'s 
too \cite{ellis}. The effective cross section entering into the 
Boltzmann equation then becomes
\begin{eqnarray}
\nonumber\sigma_{eff}&=& \sigma_{\tilde\chi\tilde\chi}
r_{\tilde\chi}r_{\tilde\chi} + 
4\sigma_{\tilde\chi\tilde\tau_2}
r_{\tilde\chi}r_{\tilde\tau_2}+ 
2(\sigma_{\tilde\tau_2\tilde\tau_2}+
\sigma_{\tilde\tau_2\tilde\tau_2^\ast})
r_{\tilde\tau_2}r_{\tilde\tau_2}+
8(\sigma_{\tilde\tau_2\tilde e_R}+
\sigma_{\tilde\tau_2\tilde e_R^\ast})
r_{\tilde\tau_2}r_{\tilde e_R}
\\ \label{sigmaeff} &+&
8 \sigma_{\tilde\chi\tilde e_R}
r_{\tilde\chi}r_{\tilde e_R}+
4(\sigma_{\tilde e_R\tilde e_R}+ 
\sigma_{\tilde e_R\tilde e_R^\ast})r_{\tilde e_R}
r_{\tilde e_R} +
4(\sigma_{\tilde e_R\tilde\mu_R}+ 
\sigma_{\tilde e_R \tilde\mu_R^\ast})r_{\tilde e_R}
r_{\tilde e_R}.
\end{eqnarray}
Here $\sigma_{ij}$ ($i,j=\tilde\chi$, $\tilde\tau_2$, 
$\tilde\tau_2^\ast$, $\tilde e_R$, $\tilde e_R^\ast$, 
$\tilde\mu_R$, $\tilde\mu_R^\ast$) is the total cross section 
for particle $i$ to annihilate with particle $j$ averaged over 
initial spin and particle-antiparticle states and the $r_i$'s can 
be found from Ref.\cite{cdm}. The Feynman graphs for 
$\sigma_{\tilde\chi\tilde\chi}$, 
$\sigma_{\tilde\chi\tilde\tau_2}$, 
$\sigma_{\tilde\tau_2\tilde\tau_2}$, 
and $\sigma_{\tilde\tau_2\tilde\tau_2^\ast}$ are listed 
in Table I of Ref.\cite{cdm}. From these diagrams, we can also 
obtain the ones for  
$\sigma_{\tilde\chi\tilde e_R}$, 
$\sigma_{\tilde e_R\tilde e_R}$, 
$\sigma_{\tilde e_R\tilde e_R^\ast}$ by replacing 
$\tilde\tau_2$ by $\tilde e_R$ 
and $\tau$ by $e$ and ignoring diagrams with $\tilde\tau_1$ 
exchange. The processes 
$\tilde\tau_2\tilde e_R\rightarrow\tau e$, 
$\tilde\tau_2\tilde e_R^\ast\rightarrow\tau\bar e$, 
$\tilde e_R\tilde\mu_R\rightarrow e\mu$ and
$\tilde e_R\tilde\mu_R^\ast\rightarrow e\bar\mu$ are 
realized via a t-channel $\tilde\chi$ exchange. The calculation 
of the $a_{ij}$'s and $b_{ij}$'s given in Ref.\cite{cdm} is readily 
extended to include these extra processes too. 

\par
The main contribution to the LSP (almost pure bino) annihilation 
cross section generally arises from stau exchange in the t- and 
u-channel leading to $\tau\bar\tau$ in the final state. 
We do not include s-channel exchange diagrams. So our results
are not valid for values of $m_{\tilde\chi}$ very close to the
poles at $m_Z/2$, $m_h/2$, $m_H/2$ or $m_A/2$ where the 
annihilation cross section is enhanced and the relic density drops 
considerably. The expressions for $a_{\tilde\chi\tilde\chi}$ 
and $b_{\tilde\chi\tilde\chi}$ can be found in Ref.\cite{cdm} 
(with the final state lepton masses neglected).

\par
The most important contribution to coannihilation arises from the 
$a_{ij}$'s. (The contribution of the $b_{ij}$'s 
($ij\neq\tilde\chi\tilde\chi$), although included in the 
calculation, is in general negligible.) The contributions of the 
various coannihilation processes to the $a_{ij}$'s and $b_{ij}$'s 
($ij\neq\tilde\chi\tilde\chi$) are calculated using techniques 
and approximations similar to the ones in Ref.\cite{cdm}. In 
particular, the contributions to the $a_{ij}$'s from the processes 
with

\begin{list}
\setlength{\rightmargin=0cm}{\leftmargin=0cm}

\item[{\bf i.}]
$\tilde\chi\tilde\tau_2$, $\tilde\tau_2\tilde\tau_2$, 
$\tilde\tau_2\tilde\tau_2^\ast$ in the initial state are 
listed in Table II of Ref.\cite{cdm}.

\item[{\bf ii.}]
$\tilde\chi\tilde e_R$, $\tilde e_R\tilde e_R^\ast$ in the 
initial state can be obtained from the formulae in Tables II and IV 
of Ref.\cite{cdm} by the replacement 
$\tilde\tau_2\rightarrow\tilde e_R$ and putting $\theta=0$, 
$m_\tau=0$.

\item[{\bf iii.}]
$\tilde\tau_2\tilde e_R$, $\tilde\tau_2\tilde e_R^\ast$, 
$\tilde e_R\tilde e_R$, $\tilde e_R\tilde \mu_R$, 
$\tilde e_R\tilde\mu_R^\ast$ in the initial state are listed 
in the following Table:

\begin{center}
TABLE. Contributions to the Coefficients $a_{ij}$
\end{center}
\begin{center}
\begin{tabular}{|c|c|}\cline{1-1} \cline{2-2}
\multicolumn{1}{|c|}{Process} & \multicolumn{1}{|c|}{Contribution
to the Coefficient $a_{ij}$}
\\ \cline{1-1} \cline{2-2}
\hline $\tilde \tau_2 \tilde e_R \rightarrow \tau e$ & $e^4 Y_R^4
\cos^2\theta m_{\tilde\chi}^2(m_{\tilde e_R}+m_{\tilde\tau_2})^2
 /$
 \\& $8 \pi c_W^4 m_{\tilde e_R} m_{\tilde\tau_2}
 (m_{\tilde\chi}^2+m_{\tilde e_R}m_{\tilde\tau_2})$
\\
\hline $\tilde \tau_2 \tilde e_R^\ast \rightarrow \tau \bar e$ &
$e^4 Y_L^2Y_R^2 \sin^2\theta m_{\tilde\chi}^2(m_{\tilde
e_R}+m_{\tilde\tau_2})^2
 /$
 \\ & $8 \pi c_W^4 m_{\tilde e_R} m_{\tilde\tau_2}
 (m_{\tilde\chi}^2+m_{\tilde e_R}m_{\tilde\tau_2})$
\\
\hline $\tilde e_R \tilde e_R \rightarrow e e$ & $e^4 Y_R^4
m_{\tilde\chi}^2 /\pi c_W^4 \Sigma_e^2$
\\
\hline $\tilde e_R \tilde \mu_R \rightarrow e \mu$ & $ e^4 Y_R^4
m_{\tilde\chi}^2/2 \pi c_W^4\Sigma_e^2$
\\
\hline $\tilde e_R \tilde \mu_R^\ast \rightarrow e \bar \mu$ & $
e^4 Y_R^4 m_{\tilde e_R}^2/12 \pi c_W^4 \Sigma_e^2$
\\
\hline
\end{tabular}
\end{center}
where $\theta$ is the stau mixing angle (not to be confused with the 
goldstino angle), $c_W=\cos\theta_W$, 
$Y_{L(R)}=-1/2(-1)$ is the hypercharge of the left(right)-handed
leptons and $\Sigma_e= m_{\tilde\chi}^2+m_{\tilde e_R}^2$ with
$m_{\tilde e_R}$ being the common mass of  $\tilde e_R$, 
$\tilde\mu_R$.
\end{list}

\par
The LSP relic abundance $\Omega_{LSP}~h^2$, which remains 
practically constant on the equispectral lines, can now be 
evaluated for any ${\rm sign}\mu$, $\tan\beta$, $m_A$ 
and $\Delta_{NLSP}$. We find that, away from the poles, 
$\Omega_{LSP}~h^2$ increases with $m_A$ (or 
$m_{\tilde\chi}$). Also, for fixed 
$m_{\tilde\chi}$, it increases with $\Delta_{NLSP}$, since  
coannihilation becomes less efficient. The mixed or the pure 
cold (in the presence of a nonzero cosmological constant) dark 
matter scenarios for large scale structure formation require  
$0.09\stackrel{_{<}}{_{\sim }}\Omega_{LSP}~h^2
\stackrel{_{<}}{_{\sim }}0.22$ \cite{cdm,lahanas}, which 
restricts $\Delta_{NLSP}$ .

\par
We will first examine the case with no Yukawa unification. As 
already mentioned, the asymptotic value of $h_b$ is specified, 
in this case, by requiring that $m_b(m_Z)$, after SUSY 
corrections, coincides with its central experimental value. 
For $\mu>0$, $m_A$ (and, thus, $m_{\tilde\chi}$) is 
forced to be quite large in order to have the 
${\rm BR}(b\rightarrow s\gamma)$ reduced below its upper 
experimental limit. Thus, the LSP and NLSP masses are required 
to be relatively close to each other so that coannihilation is 
more efficient and the bounds on $\Omega_{LSP}~h^2$ can be 
satisfied. For $\mu<0$, smaller $m_A$'s are needed for 
enhancing the $b\rightarrow s\gamma$ branching ratio so as 
to overtake its lower bound. Thus, in some regions of the 
parameter space, one can get cosmologically acceptable LSP 
relic densities even without invoking coannihilation. For 
$\mu>0$, $\tan\beta\lesssim 38$ or $\mu<0$, the Higgs 
sector turns out to be heavier than the LSP and NLSP 
($m_A\gtrsim 450~\{ 400\}~{\rm GeV}$ for $\mu>0$, 
$\tan\beta\lesssim 38$ and 
$m_A\gtrsim 340~\{ 310\}~{\rm GeV}$ 
for $\mu<0$) implying that processes with $\tau H$, $\tau A$, 
$hH$, $HH$, $H^+H^-$, $AA$ in the final state are, generally, 
kinematically blocked. Here and below, the 
limiting values obtained by including the theoretical 
uncertainty in ${\rm BR}(b\rightarrow s\gamma)$ are indicated 
in curly brackets. 

\par
\begin{figure}
\begin{center}
\epsfig{figure=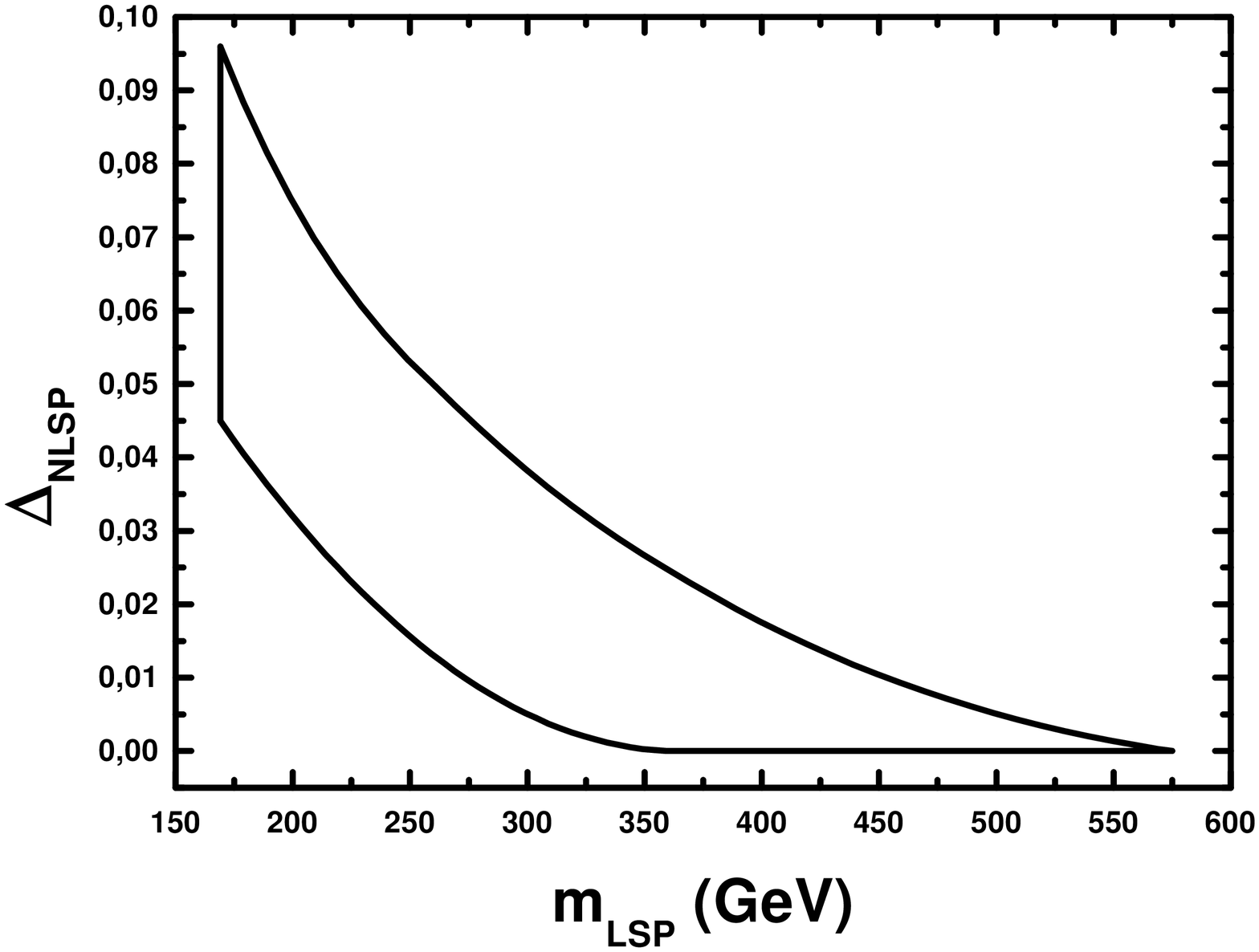,height=3.4in,angle=0}
\end{center}
\medskip
\caption{The allowed region in the $m_{LSP}-\Delta_{NLSP}$
plane for $\mu>0$ and $\tan\beta\simeq 10$ ($\epsilon=0.65$).
${\rm{BR}}(b\rightarrow s\gamma)$ is evaluated with central 
values of the input parameters and scales.
\label{deltap}}
\end{figure}

We start by constructing the regions in the 
$m_{\tilde\chi}-\Delta_{NLSP}$ plane allowed by the CDM and 
$b\rightarrow s\gamma$ considerations for each 
${\rm sign}\mu$ and $\tan\beta$. A typical example of such a 
region is shown in Fig.\ref{deltap} and corresponds to $\mu>0$ 
and $\tan\beta\simeq 10$. Here, we fixed $\epsilon=0.65$ and 
regulated $\Delta_{NLSP}$ via $\theta$. The lower bound on 
$m_{\tilde\chi}$ (almost vertical line) comes from the upper 
bound ($\simeq 4.5\times 10^{-4}$) on 
${\rm BR}(b\rightarrow s\gamma)$. The lower (upper) curved 
boundary of the allowed region corresponds to 
$\Omega_{LSP}~h^2 \simeq 0.09~(0.22)$ and the horizontal 
boundary to $\Delta_{NLSP}=0$. The maximal $m_{\tilde\chi}$ 
($\Delta_{NLSP}$) is obtained at the lower right (upper left) 
corner of this region. The value of $m_{\tilde\chi}$ can 
vary between about $169~\{ 123\}$ and $575~{\rm GeV}$. So, 
the LSP is relatively heavy and the maximal allowed 
$\Delta_{NLSP}$ is small ($\simeq 0.096~\{ 0.19\}$). 
Coannihilation is important in the whole allowed region. On the 
contrary, for $\mu<0$ and $\tan\beta\simeq 10$, we find 
lighter LSPs. Specifically, $m_{\tilde\chi}$ varies between 
about $85~\{ 79\}$ and $572~{\rm GeV}$. So, the maximal 
allowed $\Delta_{NLSP}$ is much larger 
($\simeq 0.6~\{ 0.71\}$) now, and there is a region 
($85~\{ 79\}~{\rm GeV}\lesssim m_{\tilde\chi}
\lesssim 120~{\rm GeV}$) where coannihilation is negligible. 
The lower bound on $m_{\tilde\chi}$, for $\mu<0$, corresponds 
to the lower bound on ${\rm BR}(b\rightarrow s\gamma)$ or 
$m_h$. 

\par
\begin{figure}
\begin{center}
\epsfig{figure=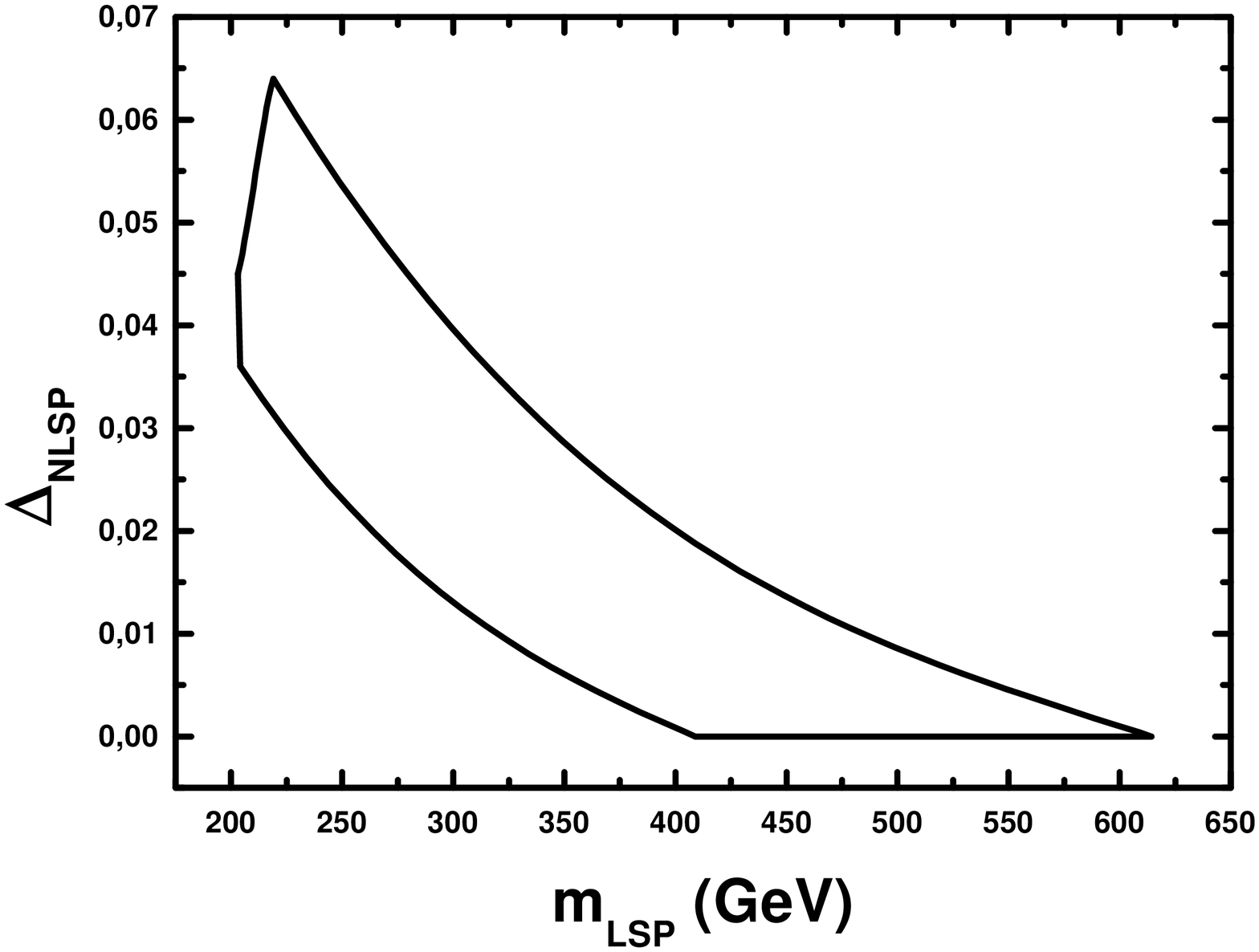,height=3.4in,angle=0}
\end{center}
\medskip
\caption{The allowed region in the $m_{LSP}-\Delta_{NLSP}$
plane for $\mu<0$, $\tan\beta\simeq 35.3$ ($\epsilon=0.99$).
${\rm{BR}}(b\rightarrow s\gamma)$ is evaluated with central 
values of the input parameters and scales. 
\label{deltac}}
\end{figure}

For $\mu<0$, there exist $\tan\beta$'s where the 
maximal $\Delta_{NLSP}$ is not obtained at the minimal 
$m_{\tilde\chi}$. This is illustrated in Fig.\ref{deltac} 
depicting the allowed region in the 
$m_{\tilde\chi}-\Delta_{NLSP}$ plane for $\mu<0$ and 
$\tan\beta\simeq 35.3$. Here, we fixed $\epsilon =0.99$. 
The LSP mass can vary between about $203$ and 
$614~{\rm GeV}$ with the lower bound corresponding to the 
lower bound on ${\rm BR}(b\rightarrow s\gamma)$. For 
the minimal $m_{\tilde\chi}$, the maximal 
$\Delta_{NLSP}$ ($\simeq 0.045$) does not 
correspond to $\Omega_{LSP}~h^2\simeq 0.22$. It is, 
rather, the absolute maximum of $\Delta_{NLSP}$ for the 
given values of ${\rm sign\mu}$, $\tan\beta$ and $m_A$ 
which is obtained at $\theta=\pi/9$ as indicated earlier 
and corresponds to $\Omega_{LSP}~h^2\simeq 0.114$. 
Increasing $m_{\tilde\chi}$, this absolute maximum of 
$\Delta_{NLSP}$ increases (along the inclined part of the left 
boundary) and $\Omega_{LSP}~h^2$ becomes $\simeq 0.22$ at 
$\Delta_{NLSP}\simeq 0.064$, which is the overall maximal 
allowed $\Delta_{NLSP}$ in this case. Including the
theoretical uncertainty in ${\rm BR}(b\rightarrow s\gamma)$,
we see that the vertical part of the boundary disappears and the 
minimal value of $m_{\tilde\chi}$ is reduced to about 
$198~{\rm{GeV}}$ corresponding to 
$\Delta_{NLSP}\simeq 0.038$.

\par
\begin{figure}
\begin{center}
\epsfig{figure=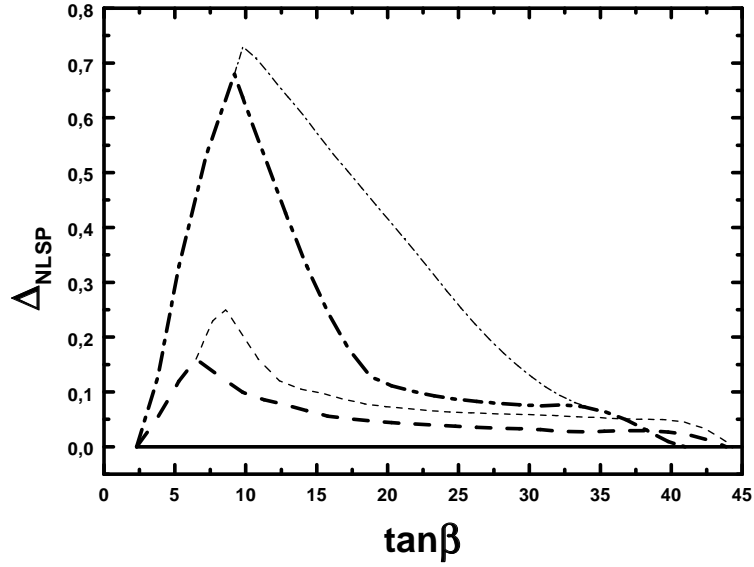,height=3.4in,angle=0}
\end{center}
\medskip
\caption{The allowed region in the $\tan\beta-\Delta_{NLSP}$ 
plane for $\mu>0~(<0)$ is between the solid and (dot-)dashed 
lines. The bold (faint) lines are without (with) 
the errors in ${\rm BR}(b\rightarrow s\gamma)$.
\label{deltam}}
\end{figure}

The maximal allowed $\Delta_{NLSP}$'s can be found for all 
possible $\tan\beta$'s and any sign of $\mu$ by repeating 
the above analysis. The results are displayed in 
Fig.\ref{deltam}, which shows the allowed regions in the 
$\tan\beta-\Delta_{NLSP}$ plane for $\mu>0$ (between the 
solid and dashed lines) and $\mu<0$ (between the solid and 
dot-dashed lines). Here, the bold (faint) lines are 
obtained by ignoring (including) the theoretical errors in 
${\rm BR}(b\rightarrow s\gamma)$, and $\epsilon$ is  
chosen for each ${\rm sign}\mu$ and $\tan\beta$ so that 
it lies in the domain of all relevant equispectral lines. We 
found that $\Delta_{NLSP}=0$ can be achieved at the maximal 
LSP mass ($\sim 600-700~{\rm GeV}$) corresponding to each 
$\tan\beta$ between 2.3 and $43.9~\{ 44.3\}$. So, the 
minimal allowed $\Delta_{NLSP}$ is always zero. Regarding 
the maximal allowed $\Delta_{NLSP}$'s, we can distinguish 
the cases:
\begin{list}
\setlength{\rightmargin=0cm}{\leftmargin=0cm}

\item[{\bf i.}]
For $\mu>0$ ($<0$) and 
$6.5~\{ 8.6\}~(9.2)\lesssim\tan\beta\lesssim 
43.9~\{ 44.3\}~(34.5)$, the maximal $\Delta_{NLSP}$ 
corresponds to the lower bound on $m_{\tilde\chi}$ found 
from the experimental limits on 
${\rm BR}(b\rightarrow s\gamma)$. The allowed regions 
are of the type in Fig.\ref{deltap} and the upper curves in 
Fig.\ref{deltam} are obtained from the upper left corners 
of these regions as we vary $\tan\beta$. For $\mu>0$, the 
lower curved boundary of the allowed regions disappears at 
high enough $\tan\beta$'s and, eventually, at 
$\tan\beta\simeq 43.9~\{ 44.3\}$, the allowed region 
shrinks to a point with 
$m_{\tilde\chi}\simeq 730~\{ 740\}~{\rm GeV}$ and 
$\Delta_{NLSP}\simeq 0$.

\item[{\bf ii.}]
For $\mu>0$ ($<0$) and $2.3\lesssim\tan\beta\lesssim 
6.5~\{ 8.6\}~(9.2~\{ 9.8\})$, the lower bound on 
$m_{\tilde\chi}$ is found from the experimental limit on 
$m_h$. This mass comes out too small for small $m_A$'s. So, 
bigger $m_A$'s (and, thus, $m_{\tilde\chi}$'s) are 
required to raise $m_h$ above $113.4~{\rm GeV}$. The allowed 
regions are again typically as in Fig.\ref{deltap} (with or 
without the curved lower boundary) and the maximal 
$\Delta_{NLSP}$ rapidly decreases with $\tan\beta$.

\item[{\bf iii.}]
For $\mu<0$ and $\tan\beta$ between about $34.5~\{ 9.8\}$ 
and 41, the maximal $\Delta_{NLSP}$ does not correspond to 
the minimal $m_{\tilde\chi}$ from the lower limit on 
${\rm BR}(b\rightarrow s\gamma)$ or $m_h$. The obtained 
allowed regions are of the type in Fig.\ref{deltac} (with or 
without the vertical part of the boundary). As $\tan\beta$ 
increases above $34.5~\{ 9.8\}$, the inclined part of their 
left boundary moves to the right and the vertical part 
eventually disappears. At even higher $\tan\beta$'s, the 
curved lower boundary also disappears and, finally, the region 
shrinks to a point at $\tan\beta\simeq 41$ with 
$\Delta_{NLSP}\simeq 0$ and 
$m_{\tilde\chi}\simeq 640~{\rm GeV}$. For low 
$\tan\beta$'s, the bino purity of the LSP decreases from above 
$98\%$ to $95\%$ and our calculation, which assumes a bino-like 
LSP, becomes less accurate. 
\end{list}
In conclusion, in the case of no Yukawa unification and for 
$\mu>0$ ($<0$), the maximal $\Delta_{NLSP}\approx 0.16
~\{ 0.25\}~(0.68~\{ 0.73\})$ is achieved at 
$\tan\beta\approx 6.5~\{ 8.6\}~(9.2~\{ 9.8\})$. Also,  
$138~\{ 114\}~(84~\{ 77\})~{\rm GeV}\lesssim 
m_{\tilde\chi}\lesssim 730~\{ 740\}~(640)~{\rm GeV}$. 
The minimal $m_{\tilde\chi}$ corresponds to the maximal 
$\Delta_{NLSP}$ except 
$m_{\tilde\chi}\approx 77~{\rm GeV}$ which is 
obtained at $\tan\beta\approx 20.4$.

\par
\begin{figure}
\begin{center}
\epsfig{figure=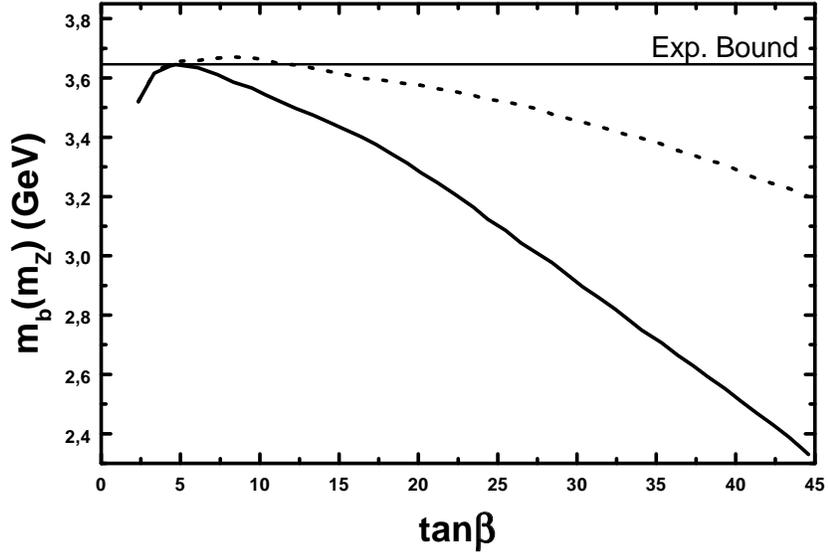,height=3.4in,angle=0}
\end{center}
\medskip
\caption{The tree-level (dotted line) and the corrected 
(solid line) $m_b(m_Z)$ versus $\tan\beta$ for 
$\Delta_{NLSP}\simeq 0$ and the minimal $m_A$. The 
experimental upper bound on $m_b(m_Z)$ is shown too.
\label{mbnew}}
\end{figure}

We now turn to the case of $b-\tau$ Yukawa unification. To 
keep $\tilde\tau_2$ heavier than $\tilde\chi$, we must 
take $\tan\beta\lesssim 45$. For $\mu<0$, the values of 
$m_b(m_Z)$, obtained from this unification assumption, turn 
out to be larger than the experimental upper limit \cite{mb} 
after including the SUSY corrections. This forces us to take 
$\mu>0$. In Fig.\ref{mbnew}, we plot the tree-level (dotted 
line) and the corrected (solid line) $m_b(m_Z)$ versus 
$\tan\beta$ for $\Delta_{NLSP}\simeq 0$ and the minimal 
value of $m_A$ which corresponds to
${\rm BR}(b\rightarrow s\gamma)\simeq 4.5\times 10^{-4}$ 
for $6.5\lesssim\tan\beta\lesssim 45$ or $m_h\approx 113.4
~{\rm GeV}$ for $2.3\lesssim\tan\beta\lesssim 6.5$. 
This choice is not crucial, because $m_b(m_Z)$, for fixed 
$\tan\beta$, turns out to be almost independent from $m_A$ 
and $\Delta_{NLSP}$. The corrected $m_b(m_Z)$ increase as 
$\tan\beta$ decreases and reaches a maximum of about 
$3.65~{\rm GeV}$ at $\tan\beta\approx 4.7$. The SUSY 
corrections decrease with $\tan\beta$. We find that, in the 
entire range $2.3\lesssim\tan\beta\lesssim 45$, the 
corrected $m_b(m_Z)$ is within the experimental limits. 

\par
Due to the relatively heavy LSP obtained with $\mu>0$, 
coannihilation is generally important for reducing 
$\Omega_{LSP}~h^2$ to an acceptable level. For 
$38\lesssim\tan\beta\lesssim 45$, the maximal allowed 
$m_{LSP}$ is raised to $\approx 790~{\rm GeV}$ due to 
the fact that the processes with $\tau H$, $\tau A$ in the 
final state are kinematically allowed. Thus, coannihilation 
is strengthened and larger $m_{LSP}$'s are allowed. On the 
contrary, for $2.3\lesssim\tan\beta\lesssim 34$, these 
processes are blocked and the upper bound on $m_{LSP}$ 
decreases to $\approx 580~{\rm GeV}$. $\Delta_{NLSP}$ 
ranges between 0 and $\approx 0.16~\{ 0.25\}$ with its 
maximum achieved at $\tan\beta\approx 6.5~\{ 8.6\}$ 
corresponding to the lowest possible 
$m_{LSP}\approx 141~\{ 115\}~{\rm GeV}$. Finally, in 
the range $34\lesssim\tan\beta\lesssim 38$, $m_{LSP}$ 
can get close to $m_A/2$, $m_H/2$ for certain $m_A$'s and 
$\Omega_{LSP}~h^2$ can be considerably reduced. Thus, the 
maximal $\Delta_{NLSP}$ and $m_{LSP}$ can be very large in 
isolated regions of the parameter space. This also applies in 
the no Yukawa unification case with $\mu>0$.

\par
In the case of $t-b$ Yukawa unification the corrected 
$m_b(m_Z)$, for $\mu<0$, again turns out to be larger than 
the experimental upper limit, so we must still choose $\mu>0$. 
We find that, for $34.3\lesssim\tan\beta$, the corrected 
$m_b(m_Z)$ is compatible with the experimental limits after 
including its theoretical uncertainties ($\approx 6\%$). 
This provides the lower 
bound on $\tan\beta$ if the theoretical uncertainties in 
${\rm BR}(b\rightarrow s\gamma)$ are included. Without 
these uncertainties, however, the lower bound on $\tan\beta$ 
is 43.7 below which the allowed region in the 
$m_{LSP}-\Delta_{NLSP}$ plane disappears. To keep 
$\tilde\tau_2$ heavier than $\tilde\chi$, we must take 
$\tan\beta\lesssim 48.5$. So there is an allowed range 
$43.7~\{ 33.5\}\lesssim\tan\beta\lesssim 48.5$ in which 
the minimal $m_{LSP}$ is about $730~\{ 507\}~{\rm GeV}$ 
with the maximal $\Delta_{NLSP}$ being 
$\approx 0~\{ 0.01\}$.

\par
For complete Yukawa unification, the lightest stau turns out to 
be lighter than the neutralino (by at least $11\%$). So, this 
case is excluded. 

\par
Theoretical errors from the implementation of the radiative 
electroweak breaking, the renormalization group analysis 
and the radiative corrections to (s)particle masses, and 
inclusion of experimental margins of various quantities can 
only further widen the allowed parameter ranges which we 
obtained. They will also produce a larger uncertainty in 
Eq.(\ref{mAc}). However, all these ambiguities are not 
expected to change our qualitative conclusions, especially 
the exclusion of complete Yukawa unification.

\par
Neutralinos could be detected via their elastic scattering with 
nuclei. For an almost pure bino, however, the cross section 
is expected to lie well below the reported sensitivity  
[$(1-10)\times 10^{-6}~\rm{Pb}$] of current experiments (DAMA). 
The reason is that the channels with Higgs and $Z$ boson (squark) 
exchange are suppressed (by the squark mass). 

\par
In summary, we studied the MSSM with radiative electroweak
breaking and boundary conditions from the Ho\v{r}ava-Witten 
theory. We assumed complete, partial or no Yukawa unification. 
The parameters were restricted by assuming that the CDM 
consists of the LSP and requiring $m_b$, after 
SUSY corrections, and ${\rm BR}(b\rightarrow s\gamma$) 
to be compatible with data. We found that complete Yukawa 
unification is excluded. Also, $t-b$ Yukawa unification is 
strongly disfavored since it requires the LSP and NLSP masses 
to be almost degenerate. This can be avoided with $b-\tau$ or 
no Yukawa unification which, for $\mu<0$, is the most natural 
case and allows the LSP mass to be as low as 
$\approx 77~{\rm GeV}$.
 
\vspace{0.5cm}

We thank M. G\'{o}mez and C. Mu\~{n}oz for discussions. S. K. 
is supported by the Spanish Ministerio de Educacion y Cultura 
and C. P. by the Greek State Scholarship Institution (I. K. Y.). 
This work was supported by the EU under TMR contract 
No. ERBFMRX--CT96--0090 and the Greek Government research grant 
PENED/95 K.A.1795.

\def\ijmp#1#2#3{{ Int. Jour. Mod. Phys. }{\bf #1~}(#2)~#3}
\def\ijmpa#1#2#3{{ Int. Jour. Mod. Phys. }{\bf A#1~}(#2)~#3}
\def\pl#1#2#3{{ Phys. Lett. }{\bf B#1~}(#2)~#3}
\def\zp#1#2#3{{ Z. Phys. }{\bf C#1~}(#2)~#3}
\def\prl#1#2#3{{ Phys. Rev. Lett. }{\bf #1~}(#2)~#3}
\def\rmp#1#2#3{{ Rev. Mod. Phys. }{\bf #1~}(#2)~#3}
\def\prep#1#2#3{{ Phys. Rep. }{\bf #1~}(#2)~#3}
\def\pr#1#2#3{{ Phys. Rev. }{\bf D#1~}(#2)~#3}
\def\np#1#2#3{{ Nucl. Phys. }{\bf B#1~}(#2)~#3}
\def\mpl#1#2#3{{ Mod. Phys. Lett. }{\bf #1~}(#2)~#3}
\def\arnps#1#2#3{{ Annu. Rev. Nucl. Part. Sci. }{\bf
#1~}(#2)~#3}
\def\sjnp#1#2#3{{ Sov. J. Nucl. Phys. }{\bf #1~}(#2)~#3}
\def\jetp#1#2#3{{ JETP Lett. }{\bf #1~}(#2)~#3}
\def\app#1#2#3{{ Acta Phys. Polon. }{\bf #1~}(#2)~#3}
\def\rnc#1#2#3{{ Riv. Nuovo Cim. }{\bf #1~}(#2)~#3}
\def\ap#1#2#3{{ Ann. Phys. }{\bf #1~}(#2)~#3}
\def\ptp#1#2#3{{ Prog. Theor. Phys. }{\bf #1~}(#2)~#3}
\def\plb#1#2#3{{ Phys. Lett. }{\bf#1B~}(#2)~#3}
\def\apjl#1#2#3{{ Astrophys. J. Lett. }{\bf #1~}(#2)~#3}
\def\n#1#2#3{{ Nature }{\bf #1~}(#2)~#3}
\def\apj#1#2#3{{ Astrophys. Journal }{\bf #1~}(#2)~#3}
\def\anj#1#2#3{{ Astron. J. }{\bf #1~}(#2)~#3}
\def\mnras#1#2#3{{ MNRAS }{\bf #1~}(#2)~#3}
\def\grg#1#2#3{{ Gen. Rel. Grav. }{\bf #1~}(#2)~#3}
\def\s#1#2#3{{ Science }{\bf #1~}(#2)~#3}
\def\baas#1#2#3{{ Bull. Am. Astron. Soc. }{\bf #1~}(#2)~#3}
\def\ibid#1#2#3{{ ibid. }{\bf #1~}(#2)~#3}
\def\JHEP#1#2#3{{ JHEP }{\bf #1~}(#2)~#3}
\def\npps#1#2#3{{ Nucl. Phys. (Proc. Sup.) }{\bf B#1~}(#2)~#3}
\def\astp#1#2#3{{ Astropart. Phys. }{\bf #1~}(#2)~#3}
\def\epj#1#2#3{{ Eur. Phys. J. }{\bf C#1~}(#2)~#3}

\end{document}